\documentclass[12pt, letterpaper]{article}

\pdfoutput=1

\usepackage{arxiv}
\usepackage[utf8]{inputenc} 
\usepackage[T1]{fontenc}    
\usepackage{amsmath,amsfonts,amssymb,bm}
\usepackage{mathtools}
\usepackage{mathptmx}
\usepackage{newtxtext}
\usepackage{newtxmath}
\usepackage{hyperref}
\usepackage{caption}
\usepackage{float}
\usepackage{xcolor}
\usepackage{multirow}
\usepackage{subcaption,booktabs}
\usepackage{rotating}
\usepackage{adjustbox}
\usepackage{graphicx}
\usepackage{array}
\usepackage{environ}
\usepackage{mathrsfs}%
\usepackage[title]{appendix}%
\usepackage{xcolor}%
\usepackage{textcomp}%
\usepackage{manyfoot}%
\usepackage{booktabs}%
\usepackage{listings}%
\usepackage{framed}
\usepackage[round]{natbib}

\title{Is a Recent Surge in Global Warming Detectable?}

\author{
 Claudie Beaulieu*(Corresponding Author) \\
Department of Ocean Sciences, \\
University of California, Santa Cruz\\
  \texttt{beaulieu@ucsc.edu} \\
  \And
Colin Gallagher\\
  School of Mathematical and Statistical Sciences\\
  Clemson University\\
  \And
Rebecca Killick \\
Department of Statistics,\\
 Lancaster University\\
  \And
Robert Lund \\
Department of Statistics,\\
University of California, Santa Cruz \\
\And
Xueheng Shi \\
Departments of Statistics and Biological Systems Engineering,\\
University of Nebraska-Lincoln \\
}

\begin{document}

\maketitle

\begin{abstract}
The global mean surface temperature is widely studied to monitor climate change. A current debate centers around whether there has been a recent (post-1970s) surge/acceleration in the warming rate.  This paper addresses whether an acceleration in the warming rate is detectable from a statistical perspective. We use changepoint models, which are statistical techniques specifically designed for identifying structural changes in time series. Four global mean surface temperature records over 1850-2023 are scrutinized within.  Our results show limited evidence for a warming surge; in most surface temperature time series, no change in the warming rate beyond the 1970s is detected. As such, we estimate minimum changes in the warming trend for a surge to be detectable in the near future.
\end{abstract}


\section{Main}
\label{sec1}

Global mean surface temperature series are crucial for monitoring global warming. The warming can be quantified by a change from a base period (e.g. pre-industrial), or by the rate of change (the warming rate) over a time interval \citep{Risbey_2017}. Considerable attention has focused on the latter in the scientific literature and media, with episodes of accelerated/decelerated warming (i.e., surges and slowdowns) being recently debated \citep{Hansen_2023,Medhaug_2017, Hausfather_2023, Mooney_Osaka_2023, Millman_2023, Harvey_2023}. 

The global mean surface temperature (GMST) naturally fluctuates in time, displaying short periods of accelerated or decelerated warming (Figure \ref{fig1}). These fluctuations may be attributed to short-term variability (or noise) in the surface temperatures. Noise in temperature series is often characterized by a short-memory process such as an autoregression. In this and other short-memory models, the ocean and other slow climate component systems respond to random atmospheric forcing slowly, producing variability at time scales longer than that of white noise \cite{Hasselman_1976}. Short-memory fluctuations can be large enough to temporarily mask a long-term warming trend, creating the appearance of a slowdown.   They can also exacerbate a warming trend, mimicking a surge \cite{Beaulieu_Killick_2018}. The key question is whether these fluctuations are occurring without any change in the underlying warming trend, or whether there has been an increase (warming surge) or decrease (warming slowdown) in the trend. To answer such questions, one needs to model the short-term variability in the GMST.  


Several studies suggested that a slowdown in warming (the so-called hiatus) occurred in the late 1990s and investigated its causes \citep{Medhaug_2017}. The slowdown was attributed to several aspects, including large-scale variability in the Pacific Ocean \cite{Kosaka_Xie_2013, Trenberth_fasullo_2013, kaufman_2011, Schmidt_2014, England_2014} and external forcings \cite{Fyfe_al_2013, Santer_2014, Schmidt_2014}. However, studies focusing on the detection of this warming pause showed that the rate of change had not declined, and that this period (from approximately 1998-2012) was not unusual given the level of short-term variability \cite{Beaulieu_Killick_2018, Cahill_2015, Lewandosky_2016, Rahmstorf_2017, Rajaratnam_2015}. More specifically, studies analyzing GMST using changepoint detection methods, which are specifically designed to objectively detect the timing of trend changes, showed no warming rate changes circa 1998 \cite{Beaulieu_Killick_2018, Cahill_2015, Rahmstorf_2017}. Further, a study assuming that the changepoint time is known and took place in 1998 showed that the trends before and after 1998 were statistically indistinguishable \cite{Rajaratnam_2015}. Overall, evidence for a pause or slowdown circa 1998 lacked a sound statistical basis. 

As per the Intergovernmental Panel on Climate Change (IPCC), detection of change refers to the "process of demonstrating that climate or a system affected by the climate has changed in some defined statistical sense without providing a reason for that change" \cite{hegerl_2010, Bindoff_2013}. Typically, the process of attribution requires that a change is statistically detected \cite{hegerl_2010}. 

The major agencies monitoring GMSTs all rank 2023 as the warmest year since the start of the instrumental record commencing in 1850 \cite{rohde_hausfather_2020, Morice_al_2021, Lenssen_2019, Vose_al_2021}. Clearly, global warming has not paused, and the current discussion about the rate of warming in the media and literature has shifted to whether there has been a warming acceleration \cite{Samset_2023, Forster_2023, Hansen_2023, Hausfather_2023, Mooney_Osaka_2023}. For example, \cite{Samset_2023} suggests that the warming rate has increased since 1990 due to a global increase in the Earth's energy imbalance (the difference in incoming solar radiation and infrared radiation emitted to space) \cite{Loeb_al_2021,vonSchuckman_al_2023}. Another recent study predicts an acceleration in the rate of warming from 2010 \cite{Hansen_2023}. With lessons learned from the hiatus debate still fresh, we assess whether a warming surge is statistically detectable. 

Changepoint techniques are used here to assess whether there has been a warming surge since the 1970s, and if so, to estimate when the surge(s) started (see Methods). Piecewise linear regression models that allow for trend changes are fitted to GMST datasets and assessed via changepoint techniques. When one does not {\em a priori} know the time of any changes, which is the case here, changepoint methods account for the number of possible different places where a new regime can begin, preventing overstating statistical significance by ``cherry-picking" the location of the changepoint times. This is a critical statistical point.  Alongside this, two types of changepoint models are used: discontinuous models where trends in the different regimes of the linear regression do not necessarily connect, and continuous models where the trend lines in successive regimes connect. While a continuous model is more physically realistic for globally averaged GMST \cite{Rahmstorf_2017}, discontinuous fits are also provided for feel; a continuous shift within a year could produce a discontinuous shift in annual measurements.  Our model fits are assessed by verifying whether the residual assumptions are met (see Supplementary Information).  Four GMST records are analyzed through 2023 at the annual time scale (see Methods). Since little evidence of a surge is concluded, a simulation study investigates how many additional years of GMST observations will be needed before a change in the warming rate will become detectable. 

\section{Results}

\subsection{Can we detect a warming surge yet?}

\begin{figure}[ht]
	\centering
	\noindent\includegraphics[width=\textwidth,angle=0]{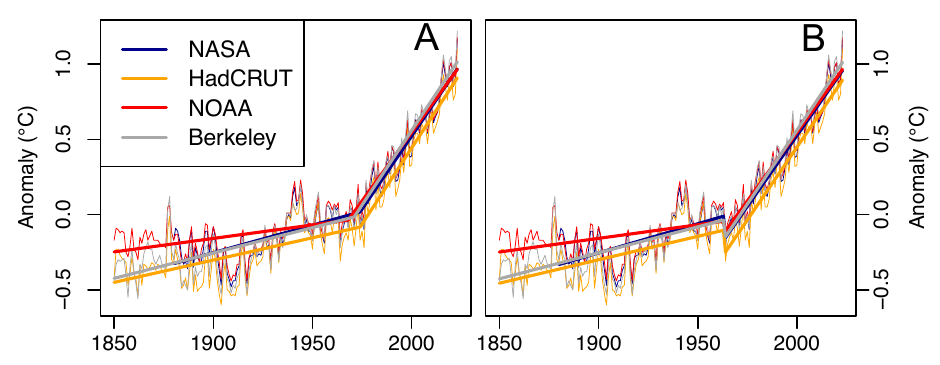}\\
	\caption{GMST anomalies from the NASA, HadCRUT, NOAA, and Berkeley series. The piecewise linear model fitted trends are superimposed for a) continuous models with changing autocorrelation, b) discontinuous models with changing autocorrelation. Note:  the model fits only show the trend in different regimes.}
	\label{fig1}
\end{figure}

Continuous and discontinuous models were fitted to all annual GMST series (see Methods). Model fits and timings of any found changepoints are listed in Table \ref{table:changepoints} and illustrated in Figure \ref{fig1}.  For the continuous model, a single changepoint is detected near 1970 in all datasets (Figure \ref{fig1}a). Similarly, we find one changepoint in all datasets within the discontinuous models (Figure \ref{fig1}b). While the timings detected are slightly earlier than those obtained with the discontinuous models, both cases do not reveal any changes in trend after the 1970s. It must be noted here that we allow the first-order autocorrelation parameter to vary between segments in the fits presented in Figure \ref{fig1}a-b. In a previous study that analyzed global surface temperature time series with changepoint detection, a reduction in autocorrelation was suggested after the 1960s \cite{Beaulieu_Killick_2018}. A changing autocorrelation allows us to better capture the larger dependence in surface temperatures in the earlier part of the record. To assess sensitivity of our results to this choice, we also include continuous and discontinuous models fitted with fixed autocorrelation parameters (same in all regimes) in the Supplementary Information. While there is variability in the number and timings of changes in the earlier part of the record, there is no warming surge detected beyond the 1970s. With a changing autocorrelation, assumptions on the residuals seem valid both for the continuous and discontinuous models, but there is leftover autocorrelation in the residuals when imposing a fixed AR(1) autocorrelation (see Supplementary Information). 

\begin{figure}[ht]
	\centering
	\noindent\includegraphics[width=\textwidth,angle=0,scale=1]{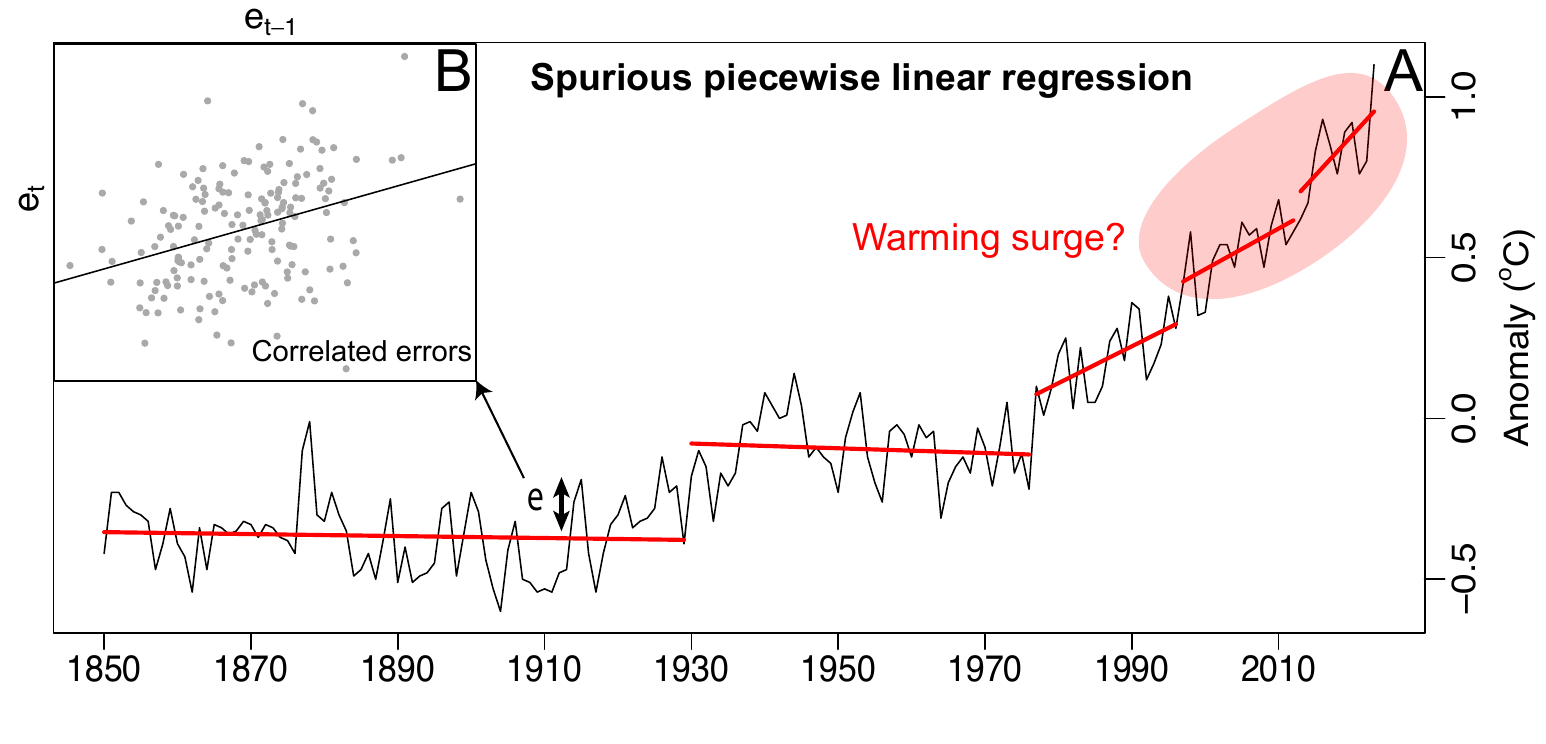}\\
	\caption{a) HadCRUT GMST anomalies with fitted superimposed discontinuous piecewise linear trends calculated assuming independent errors and b) scatterplot of the errors to illustrate a positive correlation. A hypothesis test provides strong evidence that the independent errors assumption is not valid with a $p$-value $< 0.000008$ (see Supplementary Information).}
	\label{fig2}
\end{figure}

To illustrate how a change in assumptions can yield false detections, we also fit a discontinuous model that does not take into account autocorrelation (assuming independent errors) on the HadCRUT dataset (see Figure \ref{fig2}). We include this type of model here as this the only model parameterization that yields multiple changepoints after 1970. Interestingly, multiple spurious changes in the rate of warming are detected after 1970 within this model. A warming acceleration in the 1970s is detected, followed by a different regime with a similar trend starting in 2000 and an acceleration in warming detected in 2012. Similar results are found with other datasets (see Supplementary Information). However, these fits are not valid as residuals are strongly autocorrelated in all of them (see Supplementary Information). This illustrates how changepoint analyses can produce substantially different results if autocorrelation is ignored \cite{Beaulieu_Killick_2018,Lund_al_2023,Shi_2022_CET}. Furthermore, this false detection problem is exacerbated within discontinuous fits that tend to enhance the impression of a change in trend \cite{Rahmstorf_2017}. 

\subsection{How many years are needed to detect a surge?}

While the fitted models suggest that no changepoints (surges or pauses) have occurred after the 1970s in any of the GMSTs analyzed, it would be somewhat naive to categorically conclude that no surge has occurred. In this section, we consider how far into the future a GMST must be observed to identify a statistically significant surge at the current warming rate.


With the HadCRUT GMST from 1970-2023, we computed how large a surge would need to be to become statistically detectable at the $\alpha=0.05$ significance level (see Methods). Elaborating, during the 1970-2023 period, the maximum difference in trend slopes occurs in 2012, with the estimated segments being 1970-2012 and 2013-2023, respectively.  Enforcing continuity between the two regimes, the estimated slopes are 0.019$^\circ$C/year (first segment) and 0.029$^\circ$C/year (second segment), a 53\% increase.  Accounting for the short-term variability in the HadCRUT GMST over 1970-2023 and the added uncertainty for the changepoint location, the second segment (2013-2023) would need a slope of at least 0.039$^\circ$C/year (more than a 100\% increase) to be statistically different than 0.019 at the $\alpha=0.05$ significance level right now. The estimated slope of 0.029$^\circ$C/year falls far short of this needed increase.  While it is still possible there was a change in the warming rate starting in 2013, the HadCRUT record is simply not long enough for the surge magnitude to be statistically detectable at this time.

\begin{figure}[ht]
	\centering
	\noindent\includegraphics[width=0.6\textwidth,trim={0.25cm 1cm 0.5cm 0.5cm},clip]{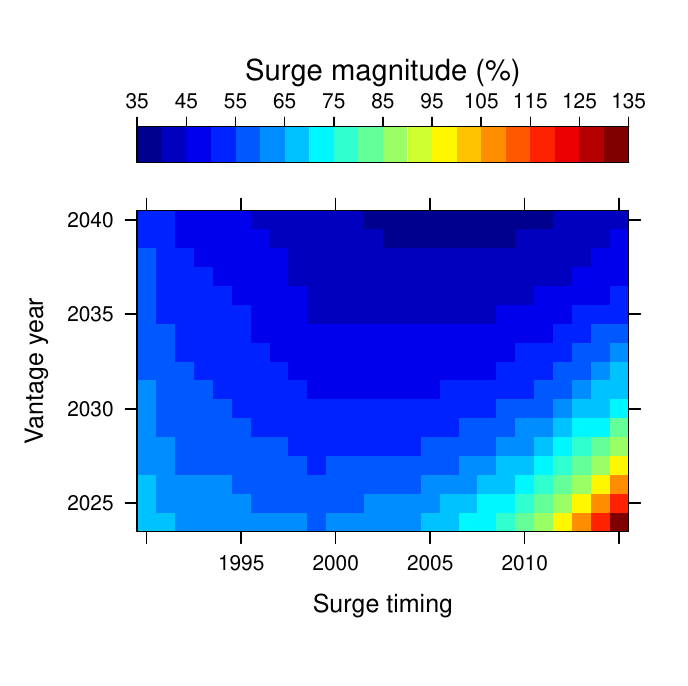}
	\caption{Minimum magnitude of a detectable (5\% critical level)  warming surge (\%) given a range of potential timings for the start of the surge and different timings for the end of the time series, based on the HadCRUT global mean surface temperature observed trend, variability and autocorrelation. Assuming a starting point in 1970, we consider a potential changepoint in trend for all years between 1990 to 2015 and potential vantage year from 2024 to 2040. The colorbar indicates the minimum magnitude of a surge to be detectable given the timing of the surge and the vantage year. }
	\label{fig3}
\end{figure}

Figure \ref{fig3} shows the magnitude of trend change required for different potential changepoint locations from 1990 to 2015 and extending the time series from 2024 until 2040. The changepoint times considered encompass surge timings suggested in the scientific literature and media \citep{Hansen_2023, Samset_2023, Hausfather_2023}. For example, to detect a warming surge that starts in 1990 over the period of 1970-2024, the magnitude of the surge needs to be at least 67\% relative to the 1970-1990 trend. This is equivalent to a change of trend from 0.018$^\circ$C/year over 1970-1990 to 0.030$^\circ$C/year over 1991-2024. If observations are extended into the future until 2030, the minimum surge detectable is 61\%, becoming 55\% by 2040. 

To detect a surge starting in 2008 as suggested in \cite{Hausfather_2023} with a 2024 vantage year, the magnitude of the surge increase needs to be at least 75\%. Extending to the vantage year 2040, a surge would need to increase at least 39\% to be detectable. 

In order to detect a warming surge starting in 2010 and ending in 2024, the trend needs to have changed by 84\% (equivalent to a trend of 0.034$^\circ$C/year from 2010-2024). With a time series extending to 2030, the surge would need change at least 58\% (a magnitude of 0.028$^\circ$C/year from 2010-2030) to be detectable. If the time series is further extended to 2040, a surge of at least 39\% change (corresponding to a magnitude of 0.026$^\circ$C/year from 2010-2040) could be detectable. 

Over the different periods in the literature, the hardest surge timing to detect is 2015 when observations end in 2024. In this case, there are only nine years of observations after the change. The trend over those nine years would need to be 133\% larger (0.044$^\circ$C/year) to become detectable. Based on Figure \ref{fig3}, it is harder to detect a surge when it occurs close to the series' end. Indeed, detection power decreases and the risk of false alarms increases with short segments \citep{Wang_2007, Beaulieu_al_2008}. 

We present this simulation based on time series properties of the HadCRUT data here however, patterns are the same across all GMST datasets (see Supplementary Information). These quotes include the added uncertainty of the changepoint location (see Methods). If the timing of a surge was known from independent observations or models, the minimum detectable surge magnitude would reduce (given the same short-term variability).



\section{Discussion}

GMST series fluctuate in time due to short-term variability, often creating the appearance of surges and slowdowns in warming. While these fluctuations may mimic an increase/decrease in the warming trend, they can simply arise from random noise in the series. This is an important consideration when discussing the warming hiatus discussion ten years ago and the more recent alleged warming acceleration. Formal detection or surges and/or pauses should account for noise (or short-term variability) and the additional uncertainty of identifying the trend change times (unless the timing of a changepoint is suggested by independent model/theory/observations).

Here, several changepoint models were used to assess whether an acceleration in warming occurred after 1970. Several different changepoint model types were considered to assess sensitivity to model choice. After accounting for short-term variability in the GMSTs (characterized by an autoregressive process), a warming surge could not be reliably detected anytime after 1970. This holds regardless of whether the changepoint models impose continuity between regimes or autocorrelation is imposed to remain fixed in time. The only way an acceleration is detected is with a model that assumes independent errors, which is not a statistically valid model choice.  
Model fits should be assessed for overall goodness of fit and produce residuals with a zero-mean and no autocorrelation (white noise). In the Supplementary Information, this is done by analyzing residuals from the model fits and testing them for residual autocorrelation.

Our analysis does not debate how unusual surface temperatures were in 2023 \citep{rantanen_2024}; our focus is on a warming rate acceleration, which is not yet detectable. However, the fact that trend changes in GMST records were not detected after the 1970s does not rule out that some small changes may have occurred; indeed, the records may be too short (or changes not large enough) to be detectable amidst the short-term variability. As such, a simulation study was conducted to assess when a warming surge will become detectable in the future. A change in the warming rate on the order of ~35\% around 2010 would become detectable circa 2035. This is the case for both an acceleration or a slowdown in warming. Our simulations allow for either an increase or decrease in the trend (two-sided test). Detection lengths would reduce with one-sided testing (say warming only), but this is not deemed justifiable given the recent discussion about a pause. Indeed, testing for a warming increase because the same observations suggest an increase will tend to overstate significance. Finally, our conclusion that an acceleration is not detectable at the global level yet may not apply at regional levels.  A focus on rigorous detection of regional warming surges should be the focus of additional work. 

Our conclusions are based on piecewise linear models.  While piecewise linear models provide a good first-order approximation of any nonlinearities and prevent us from overfitting the data, no model will perfectly describe the scenario.  The assumption that global surface temperatures contain first-order autocorrelation, which describes the dependency in the year-to-year noise values, is short-memory and geometrically decays in a year. Should the autocorrelation structure be more complex (such as long-memory where decay is a power law), there is a further risk of misinterpreting the long-memory as a trend. However, long-memory requires long series to be detectable and tends to emerge in ocean regions with strong currents \citep{Poppick_2017, Beaulieu_al_2020}.

\section{Methods}
\label{methods}

\subsection{Data}
The following four GMST time series were analyzed in this study:

\begin{itemize} 
	
	\item{The Hadley Centre/Climatic Research Unit, version 5 (HadCRUT), surface temperature \citep{Morice_al_2021}.  This series is available at \url{https://www.metoffice.gov.uk/hadobs/hadcrut5/data/current/download.html}. The annual anomalies from 1850-2023 were used. Anomalies are relative to the 1961–1990 period.}
	
	\item{The Merged Land–Ocean Surface Temperature Analysis from the National Oceanic and Atmospheric Administration (NOAA) of \cite{Vose_al_2021}.   This series is available at \url{https://www.ncei.noaa.gov/access/monitoring/climate-at-a-glance/global/time-series/globe/land\_ocean/1/9/1850-2023}. The annual anomalies from 1850-2023 were used. Anomalies are with respect to the 1971–2000 period.}
	
	\item{The Berkeley Earth Surface Temperatures (Berkeley) of \cite{rohde_hausfather_2020}.   This series is available at \url{http://berkeleyearth.org/data}. Anomalies are computed from the 1961–1990 baseline and cover the period 1850-2023.}
	
	\item{The Goddard Institute for Space Studies (GISS) Surface Temperature Analysis (GISTEMP) at the National Aeronautic Space Administration (NASA) \citep{Lenssen_2019}.   This series is available at \url{https://data.giss.nasa.gov/gistemp/} and spans 1880-2023. Anomalies are scaled to the 1951-1980 period.}
	
\end{itemize}

\subsection{Changepoint models}
Our work entails fitting several changepoint time series models that partition the GMSTs into regimes with similar trends using piecewise linear regression models. This work is most concerned with changes in the trend of the series. 

Changepoint analyses partition the data into different segments at the changepoint times.  To describe this mathematically, our model allows for $m$ changepoints during the data record $t \in \{ 1, \dots, N \}$, which occur at the times $\tau_1, \dots, \tau_m$, where the ordering $0 =\tau_0 < \tau_1 < \tau_2 < \dots < \tau_m < N=\tau_{m+1}$ is imposed.  The time $t$ segment index $r(t)$ takes the value of unity for $t \in \{ 1, \dots, \tau_1 \}$, two for $t \in \{ \tau_1 +1, \dots, \tau_2 \}, \dots$, and $m+1$ for $t \in \{ \tau_m +1 , \dots , N \}$.  Hence, the $m$ changepoint times partition the series into $m+1$ distinct segments. The model for the whole series is
\[
X_t = E[X_t] + \epsilon_t,
\]
where $E[X_t]$ is the regression function.  The regression functions considered in this manuscript include a continuous (joinpin) model, where we impose trends to meet at the changepoints, and its discontinuous counterpart.  The model errors $\{ \epsilon_t \}$ all have a zero mean and allow for autocorrelation; more about this component is said below.

The trend model regression structure we use has the simple piecewise linear form 
\[
E[X_t] = \alpha_{r(t)} + \beta_{r(t)}t,
\]
where $\beta_{r(t)}$ and $\alpha_{r(t)}$ are the trend slope and intercept, respectively, of the linear regression in force during regime $r(t)$. An equivalent representation is
\begin{equation} 
	\label{GLM}
	E[X_t] = 
	\left\{ 
	\begin{array}{cc}
		\alpha_1 + \beta_1 t,        & \qquad      0 = \tau_0  < t   \leq \tau_1,     \\
		\alpha_2 + \beta_2 t,        & \qquad \tau_1 <  t  \leq \tau_2,     \\
		\vdots                       & \vdots                               \\
		\alpha_{m+1} + \beta_{m+1}t, & \qquad \tau_m < t   \leq \tau_{m+1} = N.          \\ 
	\end{array} \right. 
\end{equation}

If continuity of the regression response $E[ X_t ]$ is imposed at the changepoint times, the restrictions 
\[
\alpha_i+\beta_i\tau_i=\alpha_{i+1}+\beta_{i+1}
\tau_i, \qquad 1 \leq i \leq m,
\]
are imposed.  These restrictions result in a model having $m$ changepoints and $m+2$ free regression parameters.  Writing the model in terms of these free parameters $\alpha_1, \beta_1, \dots, \beta_{m+1}$ gives
\[
X_t=\alpha_1 + \sum_{i=1}^{r(t)-1}(\beta_i-\beta_{i+1})\tau_i + \beta_{r(t)}t + \epsilon_t.
\]

The model errors $\{ \epsilon_t \}_ {t=1}^N$ are a zero mean autocorrelated time series that is modeled by a first-order autoregression (AR(1)).  In the global case, such a process obeys the difference equation
\begin{equation}
	\label{AR1}
	\epsilon_t = \phi \epsilon_{t-1} + Z_t,
\end{equation}
where $\{ Z_t \}$ is independent and identically distributed Gaussian noise with mean $E[ Z_t ] \equiv 0$ and variability $\mbox{Var}[Z_t] \equiv \sigma^2$, and $\phi \in (-1,1)$ is an autocorrelation parameter representing the correlation between consecutive observations in $\{ X_t \}$.  It is important to allow for autocorrelation in the model errors in climate changepoint analyses \citep{Shi_2022_CSDA}:  failure to account for autocorrelation can lead one to conclude that the estimated number of changepoints, $\hat{m}$, is larger than it should be. Higher-order autoregressions are easily accommodated should a first-order scheme be deemed insufficient.  We will also consider cases where the autoregressive parameter changes at each changepoint time; these essentially let $\phi$ depend on time $t$ via the regime index $r(t)$.   The changepoint times for the autocorrelation and mean structure are constrained to be the same.

Estimation of the model parameters proceeds as follows.  For a given number of changepoints $m$ and their occurrence times $\tau_1, \ldots , \tau_m$, one first computes maximum likelihood estimators of the regression parameters.  These produce maximum likelihood estimators of all $\alpha_i$ and $\beta_i$.  This fit gives a model likelihood, which we base on the Gaussian distribution since the series are globally and annually  averaged.   This likelihood is denoted by $L(m: \tau_1, \dots, \tau_m)$.

The hardest part of the estimation scheme lies with estimating the changepoint configuration.  This is done via a Gaussian penalized likelihood.  In particular, the penalized likelihood objective function $O$ of form
\[
O(m; \tau_1, \dots, \tau_m)= -2 \ln( L(m; \tau_1, \dots, \tau_m) ) + C(m; \tau_1, \ldots, \tau_m)
\]
is minimized over all possible changepoint configurations. The penalty $C(m; \tau_1, \dots , \tau_m)$ is a charge for having $m$ changepoints at the times $\tau_1, \dots , \tau_m$ in a model. As the number of changepoints in the model increases, the model fit becomes better and $-2\ln(L)$ correspondingly decreases.  However, eventually, adding additional changepoints to the model does little to improve its fit.  The positive penalty term counteracts ``overfitting" the number of changepoints, balancing likelihood improvements with a cost for having an excessive number of parameters within the model (changepoints + parameters within each segment).  Many penalty types have been proposed to date in the statistics literature \citep{Shi_2022_CSDA}.  One that works well in changepoint problems is the Bayesian Information Criterion (BIC) penalty 
\[
C(m; \tau_1, \dots, \tau_m) = p \ln(N),
\]
where $p$ is the total number of free parameters in the model. Table \ref{Table:penalties} lists values of $p$ for the various model types encountered in this paper.   For example, for a continuous model with a global AR(1) structure, there are $2m+4$ free regression parameters in a changepoint configuration with $m$ changepoints and $m+1$ segment parameters.   Also contributing to the parameter total are $\phi$ and $\sigma^2$.

Finding the best $m$ and $\tau_1, \dots , \tau_m$ can be accomplished via a dynamic programming algorithm called PELT \citep{Killick_2012_PELT} or a genetic algorithm search as in \cite{Davis_etal_2006, Li_2012_GA}. PELT is computationally rapid, performs an exact optimization of the penalized likelihood, and was used here.  

\subsection{Differences of Slopes Test}

How can one determine the statistical significance of a potential surge in the warming rate at some point since 1970?  To address this question, a model is needed.  Whilst our running example here considers the HadCRUT series since 1970, other GMSTs are easily analyzed.  

A simplification of (\ref{GLM}) for a single continuous changepoint is 
\begin{equation} 
	\label{GLM_1cp}
	E[X_t] = 
	\left\{ 
	\begin{array}{lc}
		\alpha_1 + \beta_1 t,                       & \qquad 0  < t   \leq k,     \\
		\alpha_1 + \beta_1 k+ \beta_2 (t-k),        & \qquad k  < t   \leq N,     \\ 
	\end{array} \right.
\end{equation}

If the timing of the single changepoint is known to be at time $k$, then the $t$-based statistic
\begin{equation}
	\label{eq:Ttest}
	T_k=\frac{\hat{\beta}_2-\hat{\beta}_1}
	{\widehat{\mbox{Var}}\left(\hat{\beta}_2-\hat{\beta}_1\right)^{\frac{1}{2}}}
\end{equation}
can be used to make inferences.  Here, $\hat{\beta}_1$ and $\hat{\beta}_2$ are the estimated trend slopes of the two segments (before and after time $k$).  One concludes a surge in warming if $T_k$ is too large to be explained by chance variation (as gauged by a $t$ distribution with $N-3$ degrees of freedom); a change in the warming rate (negative or positive) is suggested when $|T_k|$ is too large to be explained by chance variation. In computing $\mbox{Var}(\hat{\beta}_2 - \hat{\beta}_1)$, the AR(1) parameters $\phi$ and $\sigma^2$ are needed. Estimates of the two slopes, AR(1) parameters and their standard deviations are provided in time series fitting software (such as the \texttt{arima} function in \texttt{R} \citep{R}).

Extreme care is needed when $k$ is unknown. Should the time $k$ be selected visually among the many possibilities where it can occur without accounting for this, statistical mistakes can ensue. This is why changepoint techniques are needed. For example, $k=43$, which corresponds to 2012, has been suggested as a time when warming accelerated \citep{Mooney_Osaka_2023}. For the 1970-2023 data, at the 0.05 significance level, a $|T_k|$ of $2.007$ or more indicates warming rate changes; a two-tailed test was employed to allow for either an increase or a decrease in the warming rate.  For our specific example, $|T_{43}|=|0.0286-0.0187|/0.0065=1.5281$ is not statistically significant at the 0.05 significance level.


If the time of the warming rate change is unknown (as is common), statistical significance is determined based on the null hypothesis distribution of 
\begin{equation}
	\label{eq:Ttest2}
	T_{\rm max}= \max_{\ell \leq k \leq u} |T_k|,
\end{equation}
where $\ell$ and $u$ are values that truncate the admissible changepoint times for numerical stability.  The $T_{\rm max}$ statistic has significantly different statistical properties (more tail area) than $|T_k|$.  A common truncation requirement, and one that we follow, is to truncate 10\% at the series boundaries: $\ell=0.1N$ and $u=0.9N$.  If the calculated $T_{\rm max}$ statistic exceeds the threshold $Q_N$, where $Q_N$ is the 0.95 quantile of the null hypothesis distribution of $T_{\rm max}$, then a statistically significant rate change is declared with confidence 95\%.  The most likely changepoint time, $\hat{k}$, is estimated as the $k$ at which $|T_k|=T_{\rm max}$ is maximal. Statistical tests of this type are discussed in \cite{Gallagher_2013} and \cite{Robbins_2016}. There, large sample distributions were derived to determine $Q_N$.  Due to the relatively short series since 1970, we use a Monte Carlo method with Gaussian AR(1) errors to determine statistical significance.  

Elaborating, our Monte Carlo approach simulates many series using parameter estimates from the current data under the null hypothesis.  For example, with the 1970-2023 HadCRUT data, the null hypothesis parameters are estimated as $\hat{\alpha}_1=-0.17$, $\hat{\beta}_1=0.0199$, $\hat{\beta}_2=0$ (there is no second segment under the null hypothesis), $\hat{\phi}=0.0865$, and $\hat{\sigma}=0.097$.  One hundred thousand time series were then simulated, $T_{\rm max}$ was computed for each series, and 
the 0.95 quantile of these values was identified to estimate $Q_N$.  

The simulated 95th percent quantile for the HadCRUT series is $Q_N=3.1082$. The largest $|T_k|$ statistic occurs in 2012 and is $|T_{43}|=1.5281(=T_{\rm max})$, which is far from the required threshold of 3.1082.  Hence, there is little evidence for a statistically significant change in the warming rate from 1970 --- 2023 in the annual HadCRUT series; this conclusion holds for all GMST datasets considered in this paper.

So how large would the slope need to be in the second segment to declare a significant surge?  We answer this for a baseline segment of 1970-2012 and a second segment from 2013-2023. To answer this, note that $\hat{\beta}_1$ and the numerator of $T_k$ do not depend on $\hat{\beta}_2$; thus, we can set $T_{\rm max}=T_{43}=3.1082=Q_N$ and solve for $\hat{\beta}_2$.  This results in $\hat{\beta}_2 =0.0388$.  We see that a change in surge magnitude of 100(0.0388-0.0187)/0.0187 = 107\% between the two segments is required for 95\% statistical significance. 

The same logic can be used to determine future estimated rates necessary for statistical significance. Using the HadCRUT series up to 2023, there is no statistical evidence of a surge starting in 2012 relative to the 1970-2012 segment.  Will this still be the case in 2025?  How about 2040?  For any potential surge commencement year in 1990-2015, the data from 1970-2023 was used to estimate the warming trend slope, intercept, and AR(1) structure.  We then simulated cutoff quantiles for 95\% statistical significance as above for several considered vantage years, pushing out to 2040.  Since the estimated standard deviation of the slope differences depends only on the segment lengths and the AR(1) parameters, the above procedure can be solved as above for the minimal slope necessary to induce statistical significance.  

Using the HadCRUT series, one hundred thousand Gaussian series were simulated up to 2040 under our best working model (no surge, $\beta_1=0.0199$, $\alpha=-0.17$, $\phi=0.0865$, and $\sigma=0.097$).  This gives the Monte Carlo quantile estimate $Q_{71}=2.9877$.  The numerator of the $T_k$-statistic corresponding to a change starting in 2012 is estimated and solved for the minimally significant slope for the 2013-2040 segment: $\hat{\beta}_1+{\widehat{\mbox{Var}}(\hat{\beta}_1-\hat{\beta}_1)^{1/2}} Q_{71}=0.0187 + (2.9877)0.0025=0.0262$.  In short, a 40\% increase in the 2013-2040 warming rate relative to the 1970-2012 rate will be needed to declare a significant warming surge by 2040.

The above process was repeated for each commencing surge year, from 1990-2015, and each vantage year from 2024-2040.  For each surge year start $k$, the minimum statistically significant slope is calculated assuming that $T_{\rm max}=|T_{k}|$.  For each $k$, this minimally significant slope is compared to the estimated slope from the 1970-1969+$k$ series segment to calculate its associated percent change. The results for the HadCRUT series are displayed in Figure \ref{fig3}.  Overall, one sees that either significantly increased warming or many more years of observations will be required before declaring any warming sure with a reasonable degree of confidence.

\section{Tables}\label{sec5}

\begin{table}[ht]
	\caption{Changepoints detected in four global mean surface temperature datasets within continuous and discontinuous changepoint models. Trends before and after the changepoint are presented below the changepoint year (in °C/year). }
	\label{table:changepoints}
	\centering
	\begin{tabular}{c c c}
		\hline\hline 
		Dataset  & Continuous & Discontinuous \\ [0.5ex] 
		\hline
		NASA & 1973 & 1963 \\
		& 0.004, 0.020 & 0.004, 0.019 \\
		\hline
		HadCRUT & 1973 & 1963 \\
		& 0.003, 0.018 & 0.003, 0.019 \\
		\hline
		NOAA & 1967 & 1963 \\ 
		& 0.002, 0.017 & 0.001, 0.018 \\
		\hline
		Berkeley & 1970 & 1963 \\
		& 0.003, 0.019 & 0.003, 0.020 \\
		\hline
	\end{tabular}
\end{table}

\begin{table}[ht]
	\caption{Model penalties for fitting in trend models.}
	\label{Table:penalties}
	\centering
	\begin{tabular}{ccc}
		\toprule
		~ & Continuous & Discontinuous  \\
		\midrule
		Global    $\phi$  & $p=2m+4$ & $p=3m+4$ \\   
		Piecewise $\phi$  & $p=4m+4$ & $p=5m+4$ \\
		\bottomrule
	\end{tabular}
\end{table}

\begin{table}[ht]
	\caption{Estimated parameters for simulating annual global surface temperature anomalies over 1970-2023.}\label{table:parameters}%
	\centering
	\begin{tabular}{@{}lllll@{}}
		\toprule
		Dataset    & $\hat{\phi}$   &$\hat{\alpha}_1$ &$\hat{\beta}_1$& $\hat{\sigma}$   \\
		\hline
		NASA    &0.149    &-0.076  &0.019 &0.095\\
		HadCrut  &0.087    &-0.170  &0.020 &0.097\\
		NOAA     &0.190    &-0.036  &0.019 & 0.092    \\
		Berkeley &0.102    &-0.064  &0.020  &0.096     \\
		\bottomrule
	\end{tabular}
\end{table}

\section{Figures}\label{sec6}

\section{Supplement}

\subsection{Residual diagnostics for continuous and discontinuous changepoint fits}

In this section, residuals of the continuous and discontinuous models fitted in the paper are analyzed to verify our analysis assumptions.  The methods assume that the error terms $\{ \epsilon_t \}$ are normally distributed, and various time series models are implemented to account for autocorrelation in these errors.  While the statistical analyses reported in the paper are fairly robust against departures from normality, methods that do not properly account for autocorrelation can result in spurious changepoint detections \citep{Shi_2022_CSDA}.  

Portmanteau tests assess whether a fitted time series model adequately accounts for the autocorrelation in the series.  Table \ref{table:portmanteau} contains $p$-values from the Fisher-Gallagher (FG) test proposed in \cite{FisherGallagher:2012}.  Here, residual autocorrelation is jointly tested from lag 1 to lag 20.  Good fits should have no residual autocorrelation and $p$-values larger than 0.05.  The $p$-values in Table \ref{table:portmanteau} show changepoint models with independent errors contain strong evidence of residual autocorrelation.  A consequence is that these fits are likely to overestimate the number of changepoints.  A global AR(1) structure insufficiently describes autocorrelations under the continuous case, while a global AR(4) model adequately describes autocorrelations in both the continuous and discontinuous schemes.  Allowing for a changing AR(1) structure adequately models the autocorrelation in all but the Berkeley series in the continuous case.  

Normality is assessed using the Shapiro-Wilks (SW) test for the two schemes: global AR(4) and changing AR(1).  The SW test does not indicate departures from normality for any series, with all $p$-values exceeding 0.20.

\begin{table}[ht]
	\caption{Results ($p$-values) of the Fisher-Gallagher test for independence applied to residuals of the different model fits and series up to 20 lags. Bold-faced numbers indicate that the model fit is rejected due to significant residual autocorrelation.}
	\label{table:portmanteau}
	\centering
	\begin{tabular}{l c c c  c}
		\hline\hline 
		Discontinuous Fits &&&&\\
		&NASA            &HadCRUT         &NOAA     &Berkeley\\
		Independent           &{\bf 0.001} &{\bf 7e-06} &{\bf 3e-06} &{\bf 0.0005}\\
		GlobalAR(1)   &0.159  &0.304 &0.206 &{\bf 0.011}\\
		GlobalAR(4)   &0.584  &0.923 &0.611 &0.545\\
		ChangingAR(1) &0.452  &0.289 &0.232 &0.089\\
		
		\hline\hline
		Continuous Fits &&&&\\
		&NASA            &HadCRUT         &NOAA     &Berkeley\\
		Independent           &{\bf 3e-15}  &{\bf 0.003} &{\bf 5e-11} &{\bf 1e-10}\\
		GlobalAR(1)   &0.071  &{\bf 0.015} &{\bf 0.048} &{\bf 0.013}\\
		GlobalAR(4)   &0.598  &0.486 &0.535 &0.558\\
		ChangingAR(1) &0.298  &0.152 &0.161 &{\bf 0.0395}\\
		
		\hline\hline
	\end{tabular}
\end{table}

\subsubsection{Residual diagnostics for the 1970-2023 fits}
The methods in Section 4.3 are based on fits to the series since 1970. The methods used to create Figure 3 in the paper and Figures 7-9 in this supplement assume that the fitted models in Table 1 of the paper: (1) adequately describes autocorrelation, and (2) have normally distributed residuals.  Here, these two assumptions are scrutinized using the FG and SW tests. For the FG test, due to the short length of the series, we test for residual autocorrelation jointly from lag 1 to lag 10.  The $p$-values in Table \ref{table:shortfitresid} indicate that the AR(1) model adequately accounts for autocorrelation in all series.  Note that all models return no changepoints so there is no difference between the residuals from the discontinuous or continuous fits. 
There is not significant evidence of departures from normality for the NASA, Hadley and NOAA series; however, some potential departure from normality in the Berkeley series is seen although it is not significant at the 5\% level.
\begin{table}[ht]
	\caption{Results ($p$-values) of the FG test for residual autocorrelation up to 10 lags and the SW test for normality, applied to residuals of model fits to the 1970-2023 series. Note that all models return no changepoints so there is no difference between the residuals from the discontinuous or continuous fits.}
	\label{table:shortfitresid}
	\centering
	\begin{tabular}{l c c }
		\hline\hline
		Data &FG  &SW  \\
		\hline
		NASA &0.145 &0.125 \\
		Hadley &0.144 &0.107 \\
		NOAA &0.349 & 0.217\\
		Berkeley &0.157 &0.055 \\
		\hline\hline
	\end{tabular}
\end{table}

\subsection{Changepoint fits with independent errors}
\begin{figure}[ht]
	\centering
	\noindent\includegraphics[width=\textwidth,angle=0]{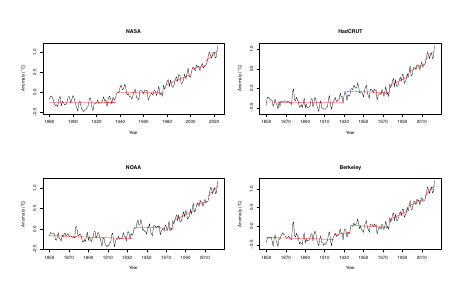}\\
	\caption{The NASA, HadCRUT, NOAA, and Berkeley series with superposed discontinuous piecewise linear regression fits assuming independent errors.  The FG test (Table \ref{table:portmanteau}) residual autocorrelation is present and these models should not be used.} 
	\label{Fig_Supp_DiscontTrend_IID}
\end{figure}

\begin{figure}[ht]
	\centering
	\noindent\includegraphics[width=\textwidth,angle=0,scale=0.5]{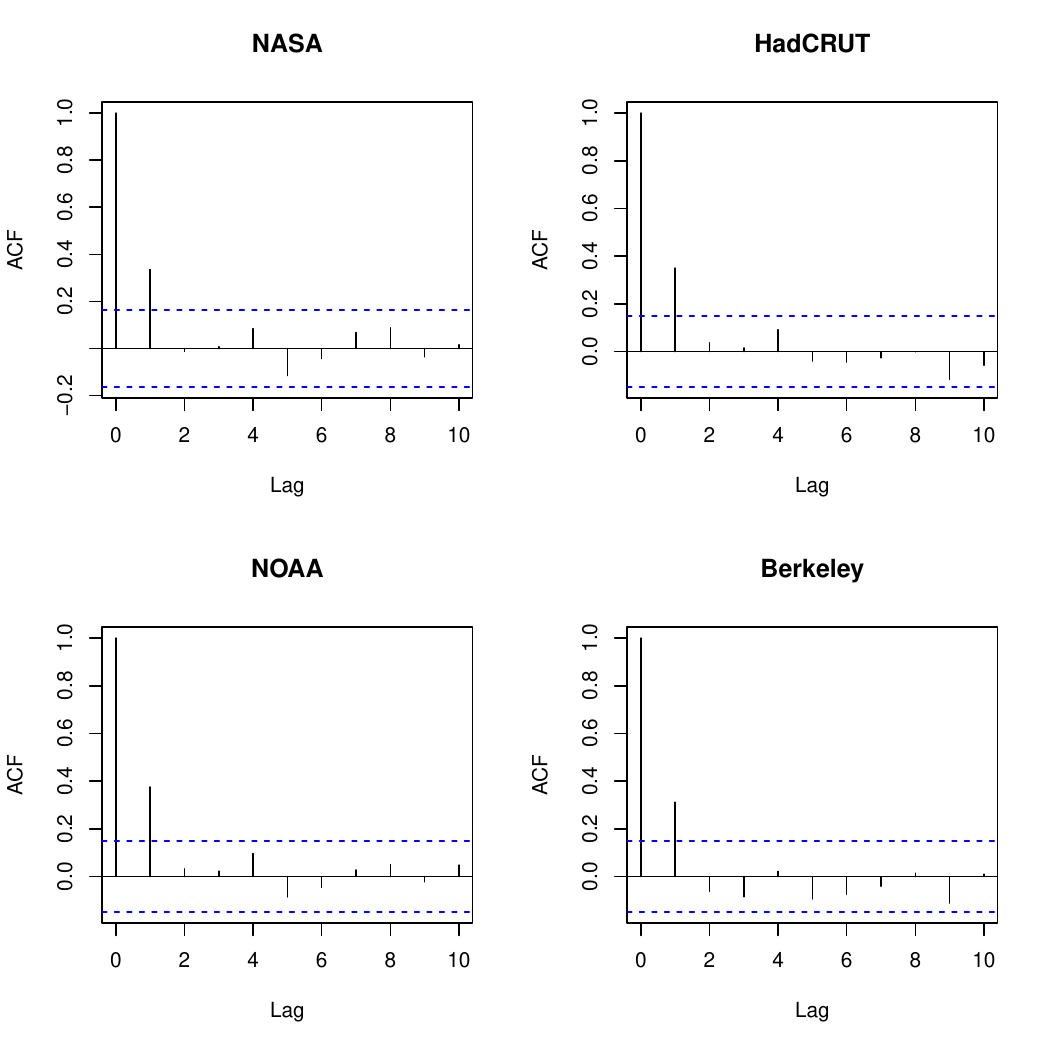}\\
	\caption{Sample autocorrelations of the residuals from the discontinuous models fitted assuming IID errors for the NASA, HadCRUT, NOAA, and Berkeley series. Horizontal dashed lines are pointwise thresholds for 95\% significance.} 
	\label{fig_discont_trend_residuals_acf}
\end{figure}

In this section, discontinuous models with independent errors are fitted to illustrate spurious model fits (Figure \ref{Fig_Supp_DiscontTrend_IID}). Analysis of the residuals from the fits with independent errors provides extremely strong evidence against independent model errors (Table \ref{table:portmanteau} and Figure \ref{fig_discont_trend_residuals_acf}), invalidating these fits.

\subsection{Changepoint fits with global autocorrelation}
\begin{figure}[ht]
	\centering
	\noindent\includegraphics[width=\textwidth,angle=0]{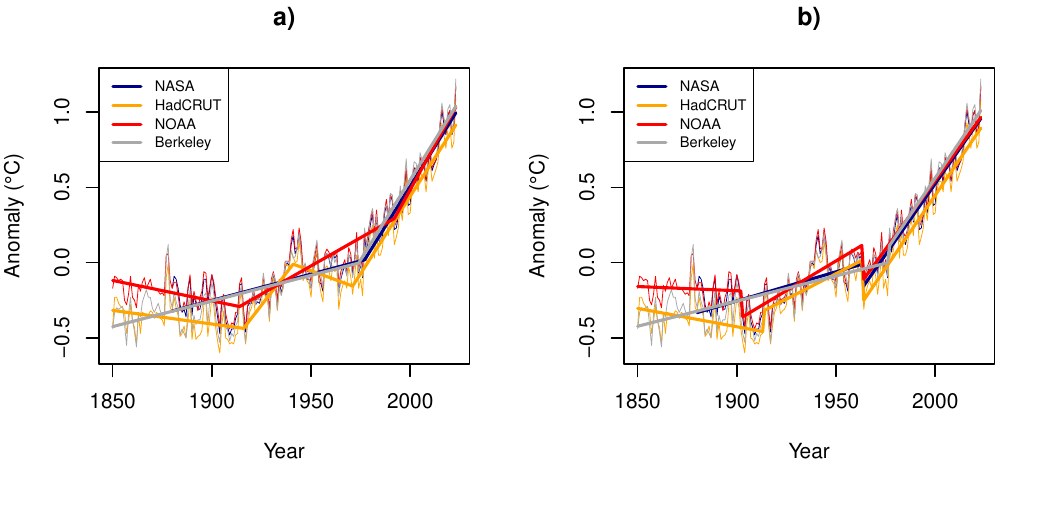}\\
	\caption{The NASA, HadCRUT, NOAA, and Berkeley series. Piecewise linear regression fits are superimposed and assume a global AR(1) structure for a) continuous models, and b) discontinuous models.}
	\label{fig_fixedARfits}
\end{figure}

For both models (continuous and discontinuous), the ``global autocorrelation" means that the first-order autocorrelation was held constant over all regimes in the fits (Figure \ref{fig_fixedARfits}). With this constant AR(1) autocorrelation structure, variability exists in the number of changepoints detected and their times in the earlier part of the GMST series. This is not surprising:  when the autocorrelation is global, correlation in the earlier part of the record is underestimated, inducing more flagged changepoints. For the discontinuous models, the global AR(1) structure sufficiently describes the autocorrelation, and changepoints occur prior to 1970 in all but the Berkeley series, which signals a final change in 1976.  Under continuity restrictions, the global AR(1) structure does not sufficiently describe series autocorrelation (Table \ref{table:portmanteau}).  

\begin{figure}[ht]
	\centering
	\noindent\includegraphics[width=\textwidth,angle=0]{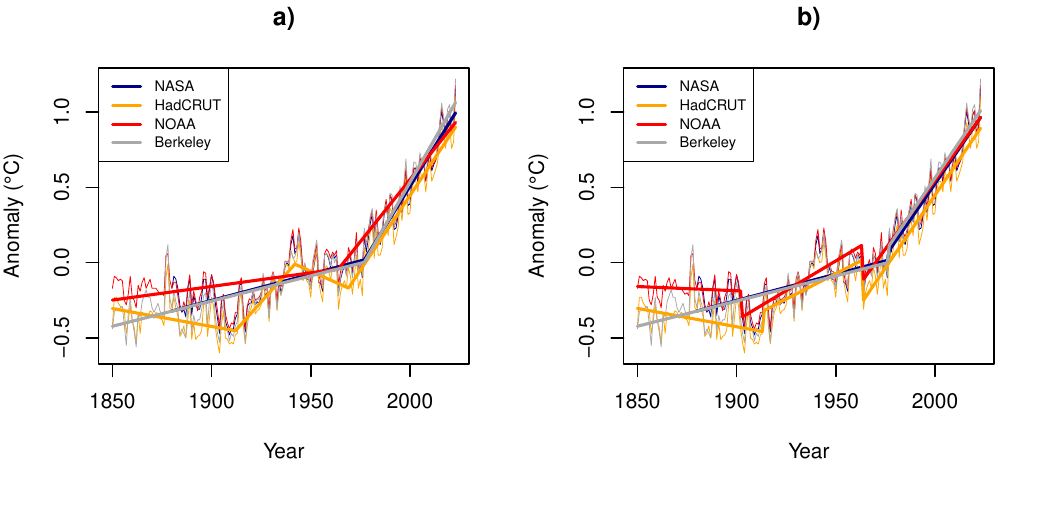}\\
	\caption{The NASA, HadCRUT, NOAA, and Berkeley series. Piecewise linear regression fits are superimposed and assume a global AR(4) structure for a) continuous models, and b) discontinuous models.}
	\label{fig_fixedAR4fits}
\end{figure}

Since the AR(1) fits were inadequate, we also fitted discontinuous and continuous models with a global AR(4) error structure (Figure \ref{fig_fixedAR4fits}). The $p$-values in Table \ref{table:portmanteau} indicate that the AR(4) model successfully describes autocorrelations in all series for both discontinuous and continuous trends. There are only minor differences in the Figure \ref{fig_fixedAR4fits} results; fewer changepoints are identified in the earlier part of the series as expected.  

\subsection{Detectability of a potential surge}

\begin{figure}[ht]
	\centering
	\noindent\includegraphics[trim=1cm 1cm 1cm 1cm, width=\textwidth,angle=0,scale=0.5]{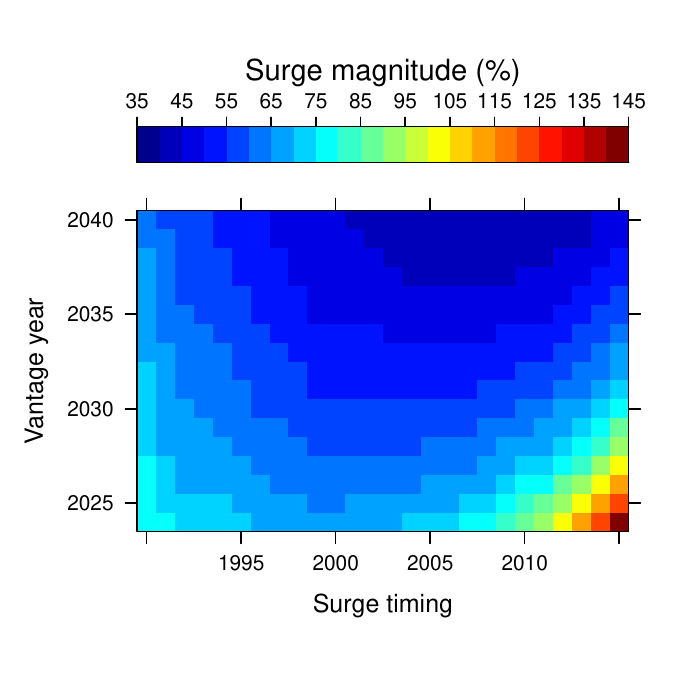}
	\caption{Minimum magnitude of a detectable (5\% critical level)  warming surge (\%), by color, based on the {\bf NASA} global mean surface temperature observed trend, variability, and autocorrelation. Assuming a starting point in 1970, we consider a potential changepoint in trend for all years between 1990 to 2015 and potential vantage year from 2024 to 2040.}
	\label{fig:NASAsims_surge}
\end{figure}

\begin{figure}[ht]
	\centering
	\noindent\includegraphics[trim=1cm 1cm 1cm 1cm, width=\textwidth,angle=0,scale=0.5]{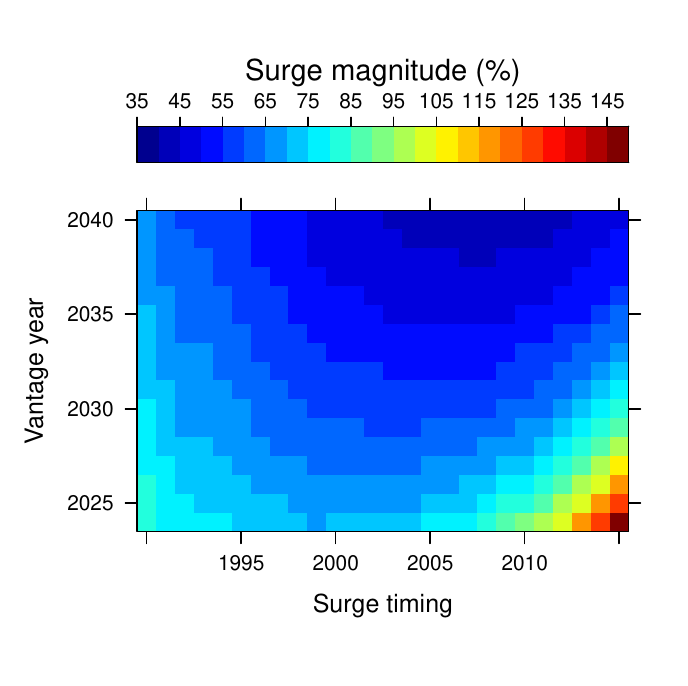}
	\caption{Minimum magnitude of a detectable (5\% critical level)  warming surge (\%), by color, based on the {\bf NOAA} global mean surface temperature observed trend, variability, and autocorrelation. Assuming a starting point in 1970, we consider a potential changepoint in trend for all years between 1990 to 2015 and potential vantage year from 2024 to 2040.}
	\label{fig:NOAAsims_surge}
\end{figure}

\begin{figure}[ht]
	\centering	\noindent\includegraphics[trim=1cm 1cm 1cm 1cm, width=\textwidth,angle=0,scale=0.5]{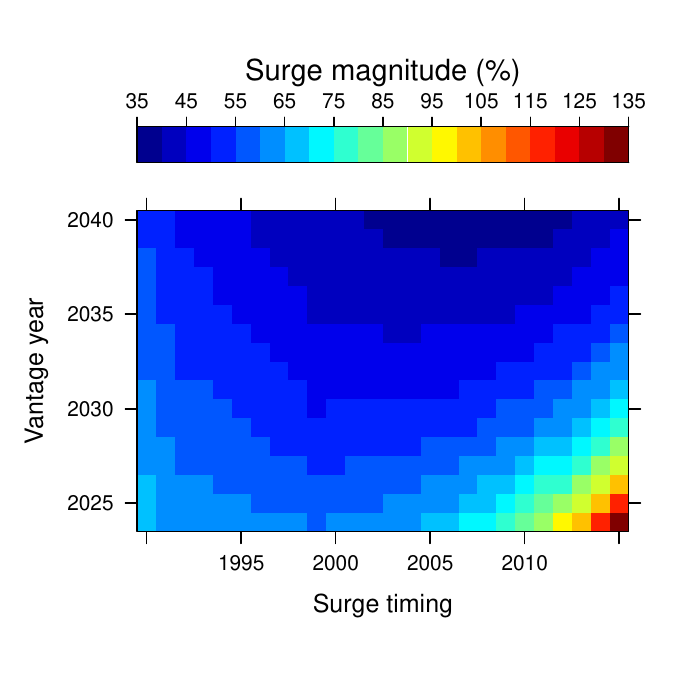}
	\caption{Minimum magnitude of a detectable (5\% critical level)  warming surge (\%), by color, based on the {\bf Berkeley} global mean surface temperature observed trend, variability, and autocorrelation. Assuming a starting point in 1970, we consider a potential changepoint in trend for all years between 1990 to 2015 and potential vantage year from 2024 to 2040. }
	\label{fig:Berkeleysims_surge}
\end{figure}

This section includes our simulation results for the minimum detectable warming surge for the NASA, NOAA, and Berkeley series (Figures \ref{fig:NASAsims_surge}, \ref{fig:NOAAsims_surge}, and \ref{fig:Berkeleysims_surge}). The HadCRUT analysis is presented in the main document.

\section*{Acknowledgement}
Rebecca Killick gratefully acknowledges funding from EP/R01860X/1 and NE/T006102/1.  Robert Lund and Xueheng Shi thank funding from NSF DMS-2113592.

%
%
\section*{Data statement}
The Central England data used in this study is available at \url{https://www.metoffice.gov.uk/hadobs/hadcet/}.  We used the annual records from 1659-2020.

\bibliographystyle{plainnat}

\end{document}